\documentstyle[aps,preprint]{revtex}

\renewcommand{\theequation}{\arabic{equation}}
\newcommand\beq{\begin{equation}}
\newcommand\eeq{\end{equation}}
\newcommand\bea{\begin{eqnarray}}
\newcommand\eea{\end{eqnarray}}
\newcommand\la{\label}
\newcommand\pa{\partial}
\newcommand\un{\underline}
\newcommand\ti{\tilde}
\newcommand\pr{\prime}
\begin{document}

\draft
\preprint{\vbox{\hbox{SOGANG-HEP 288/01}}}
\title{Improved Hamilton-Jacobi Quantization for Nonholonomic 
Constrained System}
\author{Soon-Tae Hong, Won Tae Kim, Yong-Wan Kim and Young-Jai Park}
\address{Department of Physics 
and Basic Science Research Institute,\\
Sogang University, C.P.O. Box 1142, Seoul 100-611, Korea}
\date{\today}
\maketitle
\begin{abstract}
The nonholonomic constrained system with second-class constraints is 
investigated using the Hamilton-Jacobi (HJ) quantization scheme 
to yield the complete equations of motion of the system.  Although the 
integrability conditions in the HJ scheme are equivalent to the 
involutive relations for the first-class constrained system in the improved 
Dirac quantization method (DQM), one should elaborate the HJ scheme by using 
the improved DQM in order to obtain the first-class Hamiltonian and the 
corresponding effective Lagrangian having the BRST invariant nonholonomic 
constrained system. 
\vskip 0.5cm
\par
\noindent
\end{abstract}

\section{Introduction}

Recently, several interesting constrained systems were 
investigated~\cite{park} in the framework of the improved Dirac 
quantization method (DQM)~\cite{kim}.  Based on the Carath${\acute e}$odory 
equivalent Lagrangians method~ \cite{car}, an alternative Hamilton-Jacobi 
(HJ) quantization scheme for the constrained systems was also 
proposed~\cite{g5} and this HJ scheme was exploited to 
quantize singular systems with higher order Lagrangians such as the systems 
with elements of the Berezin algebra~\cite{p11} and the Proca 
model~\cite{gu11}.  One of the most interesting application of the HJ 
quantization scheme is the systems with second-class 
constraints~\cite{dirac,gomis}, simply because the corresponding set of 
equations is not integrable~\cite{gomis}.  If the singular system is 
transformed to become completely integrable, the Hamiltonian has the form 
suitable for application of the Hamilton-Jacobi equations.

In this paper we improve the HJ quantization scheme for the nonholonomic 
constrained system (NHCS) studied in the literature~\cite{gu1,gu2} to obtain 
the complete solutions of the HJ partial differential equations (PDEs) for the 
system and to compare them with those of the standard and improved DQMs.  In 
section 2 we briefly recapitulate the HJ quantization scheme.  In section 3, 
in this refined HJ
scheme, the NHCS with nonholonomic primary constraint is reanalyzed to yield 
the complete solutions.  In section 4 we treat the NHCS by using the standard 
DQM and in section 5 we construct the first-class Hamiltonian and the 
first-class effective Lagrangian corresponding to the integrable system in 
the improved DQM.  Moreover, with the first-class effective 
Lagrangian, we have constructed the BRST invariant NHCS.  

\section{Hamilton-Jacobi quantization scheme}
\setcounter{equation}{0}
\renewcommand{\theequation}{\arabic{section}.\arabic{equation}}

In this section, we briefly recapitulate the HJ quantization 
scheme~\cite{g5,p11}.  We start with an unconstrained system with the 
Lagrangian $L$, for which we can obtain a completely equivalent 
Lagrangian described as 
\begin{equation}
L^{\prime}=L(q_i,\dot{q}_i)-\frac{dS(q_i,t)}{dt},
\label{hjact} 
\end{equation}
with $i=1,2,...,n$.  These Lagrangians are equivalent to each other if there 
exists a function $S(q_i,t)$ such that the Lagrangians $L$ and $L^{\prime}$ 
have an extreme value of the action simultaneously.  To guarantee this 
equivalence, one needs to find functions $\alpha_i(q_j,t)$ and $S(q_i,t)$ such 
that, for all neighborhood of $\dot{q}_i=\alpha_i(q_j,t)$,
\beq
L^{\prime}(q_i, \dot{q}_i =\alpha_i(q_j,t),t)=0,
\label{con1}\\
\eeq
and $L^{\prime}(q_i,\dot{q}_i)$ is positive to yield at 
$\dot{q}_i=\alpha_i(q_j,t)$,
\beq
\frac{\partial L^{\prime}}{\partial {\dot q}_{i}}=0.
\label{dotq}
\eeq
Note that the Lagrangian $L^{\prime}$ has now 
a minimum at $\dot{q}_i=\alpha_i(q_j,t)$ so that the solutions of the 
differential equations given by $\dot{q}_i=\alpha_i(q_j,t)$ can yield the 
extremal action.  Now exploiting Eqs. (\ref{hjact}) and (\ref{con1}) one can 
obtain at $\dot{q}_{i}=\alpha_{i}$ 
\begin{equation}
\label{con11}
\frac{\partial S}{\partial t}=L-\frac{\partial S}{\partial q_i}\dot{q}_{i}.
\end{equation}
Similarly, combining Eq. (\ref{hjact}) and (\ref{dotq}) yields at 
$\dot{q}_{i}=\alpha_{i}$ the HJ equation 
\beq
\frac{\partial S}{\partial q_i}=p_{i}, 
\la{con22}
\eeq
where $p_{i}$ are the conjugate momenta.  Inserting $p_{i}$ into 
Eq. (\ref{con11}), one can obtain the HJ PDE in terms of the Hamiltonian 
$H_{0}$ as follows
\begin{equation}
\frac{\partial S}{\partial t}=-H_0=-p_i\dot{q}_i+L(q_i,\dot{q}_i).
\label{canham}
\end{equation}

Next, we consider a constrained system in which canonical variables are 
not all independent.  In the constrained system, the Lagrangian $L$ is 
singular so that the determinant of the Hessian matrix 
$H_{ij}=\frac{\partial^2 L}{\partial\dot{q}_i\partial\dot{q}_j}$ is zero 
and the accelerations of some variables $\ddot{q}_i$ are not 
uniquely determined by the positions and the velocities at a given time. 

Now we consider the rank $n-m$ of the Hessian where determinant of a 
sub-matrix of the Hessian is not zero and thus some velocities 
$\dot{q}_a$ ($a=1,2,..., n-m$) can be solved as a function of coordinates 
$q_i$ and momenta $p_a$ to yield $\dot{q}_a=\dot{q}_a (q_i,p_b)$.  The 
remaining momenta $p_\alpha$ ($\alpha=n-m+1,...,n$) are functions of $q_i$ 
and $p_a$ to yield 
\begin{equation}
p_\alpha = -H_\alpha(q_i,p_a),
\la{palpha}
\end{equation} 
which are equivalent to the primary constraints $p_\alpha+H_\alpha$ in the 
Dirac terminology~\cite{dirac}.  The Hamiltonian (\ref{canham}) then becomes
\begin{equation}
\label{hjham}
H_0=p_a\dot{q}_a+p_\alpha\dot{q}_\alpha-L(q_i,\dot{q}_{a},\dot{q}_\alpha),
\end{equation}
which can be shown not to depend explicitly on the velocities 
$\dot{q}_\alpha$.

With the redefinition: $t_{\un{\alpha}}=(t_{0},t_{\alpha})=(t,q_{\alpha})$, 
($\un{\alpha}=0,n-m+1,...,n$) and $p_{0}=\frac{\partial S}{\partial t}$,  
Eqs. (\ref{canham}) and (\ref{palpha}) yield the generalized HJ PDEs 
for $\un{\alpha}=0,n-m+1,...,n$,
\beq
\label{hjpde}
H^{\prime}_{\un{\alpha}} \equiv p_{\un{\alpha}}+H_{\un{\alpha}}
(t_{\un{\beta}},q_a,p_a)=0.
\eeq
Exploiting Eqs. (\ref{hjham}) and (\ref{hjpde}), one can obtain 
\begin{equation}
\label{hjem}
dq_{\un{i}} =\frac{\partial H^{\prime}_{\un{\alpha}}}{\partial p_{\un{i}}}dt_{\un{\alpha}},~~~
dp_{\un{i}}=-\frac{\partial H^{\prime}_{\un{\alpha}}}
{\partial q_{\un{i}}}dt_{\un{\alpha}}.
\end{equation}
Here note that we have used the extended index $\un{i}$ ($\un{i}=0,1,...,n$), 
instead of the index $a$ ($a=1,2,...,n-m$) used in the literature~\cite{gu1}, 
to obtain the complete solutions to the system.  Eq. (\ref{hjem}) then 
yields
\begin{equation}
\label{hjem1}
dS=\left(-H_{\un{\alpha}}+p_a\frac{\partial H^{\prime}_{\un{\alpha}}}{\partial p_a}\right)
dt_{\un{\alpha}},
\end{equation}
from which we obtain the action of the form
\beq
S=\int (-H_{0}dt+p_{i}dq_{i}).
\la{action0ii}
\eeq
Note that at the moment $dq_{i}$ cannot be integrable to yield the 
desired effective Lagrangian, which will be realized by introducing 
auxiliary fields in the improved DQM in the next sections.    

Now, in order to discuss the integrability conditions, one can introduce a
linear operator $X_{\un{\alpha}}$ ($\un{\alpha}=0,n-m+1,...,n$) corresponding 
to Eq. (\ref{hjem}) as
\begin{equation}
\label{partial}
X_{\un{\alpha}} f= \{f, H^{\prime}_{\un{\alpha}} \}
=\frac{\pa f}{\pa q_{\un{i}}}\frac{\pa H^{\prime}_{\un{\alpha}}}{\pa p_{\un{i}}}
-\frac{\pa H^{\prime}_{\un{\alpha}}}{\pa q_{\un{i}}}\frac{\pa f}
{\pa p_{\un{i}}},
\end{equation}
from which one can obtain the bracket relations among the linear operators 
$X_{\un{\alpha}}$ 
\begin{equation}
[X_{\un{\beta}}, X_{\un{\alpha}}]f=\{f, \{H^{\prime}_{\un{\alpha}}, 
H^{\prime}_{\un{\beta}}\}\}.
\la{comm}
\end{equation}
Note that, if one can introduce operators $X_{\bar{\alpha}}$ with an extended 
index $\bar{\alpha}$ ($\bar{\alpha}=0,n-m+1,...,n,...$) such that these 
operators satisfy a 
closed Lie algebra 
\beq
[X_{\un{\beta}}, X_{\bar{\alpha}}]f=\{f, \{H^{\prime}_{\bar{\alpha}}, 
H^{\prime}_{\un{\beta}}\}\}=0,
\la{comm2}
\eeq
then the system of PDEs $X_{\un{\beta}}f=0$ is {\it complete} and the total 
differential equations $dq_i=f_{i\un{\beta}}dt_{\un{\beta}}$ is called 
{\it integrable}.  Since the total differential for any function 
$F$ can be written as $dF=\{F,H_{\un{\beta}}^{\prime}\}dt_{\un{\beta}}$, the 
integrability conditions for $\bar{\alpha}=0,n-m+1,...,n,...$, are given as 
\begin{equation}
\dot{H}^{\prime}_{\bar{\alpha}}=\{H^{\prime}_{\bar{\alpha}},H^{\prime}_{0}\}
+\{H^{\prime}_{\bar{\alpha}},H^{\prime}_{\beta}\}\dot{q}_{\beta}=0.
\la{integ2}
\end{equation}
Note that the definition of the brackets (whose index $\un{i}$ runs from 0 to 
$n$) in Eq. (\ref{integ2}) slightly differs from that of usual Poisson 
brackets (whose index $i$ runs from 1 to $n$).  If $H_{\beta}^{\prime}$ does 
not possess time-dependence explicitly, the integrability conditions 
(\ref{integ2}) are then equivalent to the consistency conditions in the DQM 
and the involution relations in the improved DQM, which will be discussed 
in the next sections.

\section{NHCS in Hamilton-Jacobi scheme}
\setcounter{equation}{0}
\renewcommand{\theequation}{\arabic{section}.\arabic{equation}}

In this section, we consider the {\it nonholonomic} constrained system (NHCS), 
where the primary constraint cannot be expressed in terms of the coordinates 
only, by introducing the Lagrangian of the form~\cite{gu1,gu2}
\begin{equation}
\label{action}
L_{0} = \frac{1}{2} \dot{q}^2_1 - \frac{1}{4}(\dot{q}_2 - \dot{q}_3)^2 + (q_1+q_3)\dot{q}_2
    -q_1-q_2-q^2_3,
\end{equation}
with the canonical momenta 
\beq
\label{momenta}
p_1 =\dot{q_{1}},~~~
p_2 =\frac{1}{2}(\dot{q}_3-\dot{q}_2)+q_1+q_3,~~~
p_3 =\frac{1}{2}(\dot{q_{2}}-\dot{q_{3}}).
\eeq
Since the rank of the Hessian matrix $H_{ij}$ ($i,j=1,2,3)$ is two, we have 
two independent relations of the momenta $p_{1}$ and $p_{3}$, and the 
dependent one $p_{2}$ given as
\begin{equation}
\label{primary}
p_2=-p_3+q_1+q_3 =-H_{2},
\end{equation}
which is a {\it nonholonomic} primary constraint in the Dirac 
terminology~\cite{dirac}.

On the other hand, the Hamiltonian given as
\begin{equation}
\label{canH}
H_0 = \frac{1}{2}(p^2_1-2p^2_3)+q_1+q_2+q^2_3,
\end{equation}
and Eqs. (\ref{hjpde}) and (\ref{primary}) yield the generalized HJ PDEs for 
$H^{\prime}_{\un{\alpha}}$ ($\un{\alpha}=0,2$)
\beq
H'_0 = p_0+H_{0}=0,~~~
H'_2 = p_2+p_3-q_1-q_3 =0.
\la{hprime2}
\eeq
Since the Hamilton equations are given by Eq. (\ref{hjem}), the above
$H^{\prime}_{\un{\alpha}}$ $(\un{\alpha}=0,2)$ functions generate the 
following set of equations of motion
\beq
\begin{array}{llll}
dq_0=dt, &dq_1=p_1dt,    &dq_2=dq_2, &dq_3=-2p_3dt+dq_2,\\
dp_0=0,  &dp_1=-dt+dq_2, &dp_2=-dt,  &dp_3=-2q_3dt+dq_2.
\end{array}
\label{eq}
\eeq
Note that, since $dq_2$ is trivial, one could not obtain any information at 
this level.

Next, for the above $H'_0$ and $H'_2$, the integrability 
conditions (\ref{integ2}) then imply
\begin{equation}
\dot{H}^{\prime}_0 =\{H_{0}^{\prime},H_{2}^{\prime}\}\dot{q}_{2}
=-H_{3}^{\prime}\dot{q}_{2}=0,~~~
\dot{H}^{\prime}_2 =\{H_{2}^{\prime},H_{0}^{\prime}\}=H_{3}^{\prime}=0,
\la{integ3}
\end{equation}
with $H_{3}^{\prime}$ being a nonholonomic secondary constraint of the form 
\begin{equation}
H'_3 =2p_3-p_1-2q_3-1.
\la{hprime3}
\end{equation}
This $H_{3}^{\prime}$ then yields an additional integrability condition 
\beq
\dot{H}^{\prime}_3 =\{H_{3}^{\prime},H_{0}^{\prime}\}
                   +\{H_{3}^{\prime},H_{2}^{\prime}\}\dot{q}_{2}
                   =4p_3-4q_3+1-\dot{q}_2=0,
\la{hp3}
\eeq
to arrive at the desired information of $dq_2$ 
absent in Eq. (\ref{eq}),
\begin{equation}
\label{q2eq}
\ddot{q}_2-2\dot{q}_2+2=0,
\end{equation}
so that we can now solve the equations of motion completely.  Here one 
notes that using Eq. (\ref{momenta}) the nonholonomic constraint 
(\ref{hprime3}) can be rewritten in terms of the $q_{i}$ and $\dot{q}_{i}$ as 
below
\beq
H^{\prime}_{3}=-\dot{q}_{1}+\dot{q}_{2}-\dot{q}_{3}-2q_{3}-1.
\la{nhconst0}
\eeq 

With the aid of Eq. (\ref{q2eq}), we can now completely find the solutions 
for the equations of motion (\ref{eq}) as
\beq
\begin{array}{ll}
q_1(t)=Ae^{2t}-t+C_1,
& p_1(t)=2Ae^{2t}-1,\\
q_2(t)=2Ae^{2t}+t+C_2,
& p_2(t)=-t+C_1,\\
q_3(t)=\frac{A}{2}e^{2t}+Be^{-2t}+\frac{1}{2}, 
& p_3(t)=\frac{3}{2}Ae^{2t}+Be^{-2t}+\frac{1}{2},
\end{array}
\label{sol}
\eeq
where $A$, $B$, $C_1$ and $C_2$ are arbitrary constants of 
integration.  Note that the results (\ref{sol}) are exactly same as 
those of Ref.~\cite{gu1,gu2} except the existence of $C_1$ in our solutions.

\section{NHCS in standard Dirac quantization method}
\setcounter{equation}{0}
\renewcommand{\theequation}{\arabic{section}.\arabic{equation}}

In this section, we analyze the Hamiltonian structure of the Lagrangian 
(\ref{action}) in the standard DQM to compare with that in 
the HJ scheme.  With the definition of the canonical momenta (\ref{momenta}) 
one can obtain the {\it nonholonomic} primary constraint of the form, which 
is the same as $H^{\prime}_{2}$ defined in Eq. (\ref{hprime2}),  
\beq
\Omega_1 =p_2+p_3-q_1-q_3 \approx 0.
\la{const1}
\eeq

Now we define the Hamiltonian $H$ with a Lagrangian multiplier $v$, 
\beq
H=H_{0}+v\Omega_{1},
\la{hc}
\eeq
from which, requiring the time stability of the primary constraint 
(\ref{const1}), one can easily find secondary constraint, which is equal to 
$H^{\prime}_{3}$ in Eq. (\ref{hprime3}), 
\beq
\dot{\Omega}_{1}=\{\Omega_{1},H\}=\Omega_2=2p_3-p_1-2q_3-1\approx 0.
\la{const2}
\eeq
The time stability of $\Omega_{2}$ yields 
\beq
\dot{\Omega}_{2}=\{\Omega_{2},H\}=4p_{3}-4q_{3}+1-v=0,
\la{const3}
\eeq
to fix the $v$ as $v=4p_{3}-4q_{3}+1$.  These constraints in 
Eqs. (\ref{const1}) and (\ref{const2}) make the system 
second-class with $\Delta_{ab}=\{\Omega_a, \Omega_b\}=\epsilon_{ab}$ and 
$\epsilon_{12}=1$.  

Next, to obtain the equations of motion for $q_{i}$ $(i=1,2,3)$, we can 
proceed to construct the Poisson brackets 
$\{q_{i},H\}$ and $\{p_{i},H\}$ for the physical variables 
$(q_{i},p_{i})$ to yield
\beq
\begin{array}{lll}
\dot{q}_{1}=q_{1}, 
& \dot{q}_{2}=4p_{3}-4q_{3}+1, 
&\dot{q}_{3}=4p_{2}+6p_{3}-4q_{1}-8q_{3}+1,\\
\dot{p}_{1}=4p_{3}-4q_{3}, 
& \dot{p}_{2}=-1, 
&\dot{p}_{3}=4p_{2}+8p_{3}-4q_{1}-10q_{3}+1,
\end{array}
\la{pqdots}
\eeq
which, together with the constraints (\ref{const1}) and (\ref{const2}), 
reproduce the solutions (\ref{sol}).  Note that the equation of 
motion for $\dot{q}_{2}$ in Eq. (\ref{pqdots}) is the same as the 
above fixed value of $v$.  Note that the Poisson brackets in the HJ scheme are 
the same as those in the DQM since $\Omega_{a}$ do not depend on time 
explicitly.  Moreover, if the integrability conditions (\ref{integ3}) and 
(\ref{hp3}) are rewritten in terms of $\Omega_{1}(=H_{2}^{\prime})$ and 
$\Omega_{2}(=H_{3}^{\prime})$ and $v(=\dot{q}_{2})$, one can easily reproduce 
Eqs. (\ref{const2}) and (\ref{const3}) to explicitly show that the 
integrability conditions in HJ scheme is equivalent to the consistency 
conditions in DQM.

On the other hand, in order to consistently quantize the NHCS, one 
should obtain the Dirac brackets 
\begin{eqnarray}
\label{dirac-a}
&& \{q_i, p_j\}_D = \delta_{ij}-\delta_{i1}(\delta_{j1}+\delta_{j3})
                  -2\delta_{i2}\delta_{j3}+2\delta_{i3}\delta_{j1},
                  \nonumber \\
&& \{q_i, q_j\}_D = -2(\delta_{i2}\delta_{j3}-\delta_{i3}\delta_{j2})
                    - (\delta_{i1}\delta_{j2}-\delta_{i2}\delta_{j1})
                    + (\delta_{i3}\delta_{j1}-\delta_{i1}\delta_{j3}),
                  \nonumber \\
&& \{p_i, p_j\}_D = -2(\delta_{i1}\delta_{j3}-\delta_{i3}\delta_{j1}),
\end{eqnarray}
where the Dirac brackets for any functions $A(q,p), B(q,p)$ are defined as 
$\{A, B\}_D = \{A, B\} - \{A, \Omega_a\}\Delta^{ab}\{\Omega_b, B\}$, with 
$\Delta^{ab}$ being the inverse of $\Delta_{ab}$.

\section{NHCS in improved Dirac quantization method}
\setcounter{equation}{0}
\renewcommand{\theequation}{\arabic{section}.\arabic{equation}}
\subsection{Gauge invariant Lagrangian for NHCS}

Now, according to the improved DQM~\cite{park,kim}, we embed the second-class 
constrained system into first-class one via systematic first-class 
prescription, where one introduces an auxiliary canonical pairs 
$(\theta, \pi_\theta)$ satisfying  $\{\theta,\pi_{\theta}\}=1$ to yield 
modified first-class constraints $\ti{\Omega}_a$ satisfying a closed 
Lie algebra $\{\ti{\Omega}_a, \ti{\Omega}_b \} =0$.  Following the 
improved DQM~\cite{park,kim} we can find effective first-class 
constraints as
\beq
\label{effcon}
\ti{\Omega}_1 = \Omega_1 + \theta,~~~
\ti{\Omega}_2 = \Omega_2 - \pi_\theta, 
\eeq
whose Poisson brackets strongly vanish in the extended phase space due to 
the introduction of the auxiliary canonical pairs of $(\theta, \pi_\theta)$.
Note that, in the limit $(\theta,\pi_{\theta})\rightarrow 0$, 
$\ti{\Omega}_{1}$ and $\ti{\Omega}_{2}$ reduce into $H_{2}^{\prime}$ and 
$H_{3}^{\prime}$ in Eqs. (\ref{hprime2}) and (\ref{hprime3}) of the HJ scheme, 
respectively.  On the other hand, we can also obtain first-class physical 
variable in this first-class embedded phase space as
\beq
\begin{array}{lll}
\ti{q}_1 = q_1-\theta, &\ti{q}_2= q_2+\pi_\theta,
&\ti{q}_3 = q_3+\pi_\theta+2\theta,\\
\ti{p}_1 = p_1+\pi_\theta, &\ti{p}_2= p_2,
&\ti{p}_3 = p_3+\pi_\theta+2\theta.
\end{array}
\label{phyvar}
\eeq
Note that the Dirac algebra (\ref{dirac-a}) in the original phase space
is mapped into the Poisson algebra of the first-class physical variables in 
the extended phase space.  Moreover, since the second-class nature 
of the NHCS is sometimes suffering from unfavorable problems such as 
ordering upon quantization, it is preferred to convert the second-class 
system into the first-class one so that one can perform consistent 
quantization.  Using the first-class physical variables (\ref{phyvar}), we 
can now construct the first-class Hamiltonian as
\begin{equation}
\label{firstH}
\ti{H}=H_0+(-4p_3+4q_3-1)\theta +(p_{1}-2p_{3}+2q_{3}+1)\pi_{\theta}
+\frac{1}{2}\pi^2_{\theta},
\end{equation}
where $H_0$ is the original Hamiltonian (\ref{canH}).  Moreover, one can also 
construct the equivalent first-class Hamiltonian of the form
\beq
\ti{H}^{\prime}=\ti{H}+\pi_{\theta}\ti{\Omega}_{2},
\la{htp}
\eeq
to satisfy the Gauss law constraints 
$\{\ti{\Omega}_{1},\ti{H}^{\prime}\}=\ti{\Omega}_{2}$ and 
$\{\ti{\Omega}_{2},\ti{H}^{\prime}\}=0$.  Note that these constraint Lie 
algebra are the same as those in the HJ scheme in the 
$(\theta,\pi_{\theta})\rightarrow 0$ limit to show that the involution 
relations are equivalent to the integrability conditions in the HJ scheme in 
this limit, as in the standard DQM.

Now, exploiting the gauge invariant first-class effective Hamiltonian, we 
perform the Legendre transformation to integrate out all the involved momenta, 
via the partition function 
\begin{equation}
Z=\int \prod_{i=1}^{3}dq_{i}dp_{i}d\theta d\pi_\theta \prod_{a,b=1}^{2}
\delta(\ti{\Omega}_a)\delta(\Gamma_b) 
{\rm det}\mid \{\ti{\Omega}_a,\Gamma_b\}\mid  e^{i\int dt~L},
\end{equation}
where $\Gamma_a$ $(a=1,2)$ is a gauge-fixing function and the
effective Lagrangian is given as
\begin{equation}
L=p_i\dot{q}_i+\pi_\theta\dot{\theta}-\ti{H}^{\pr}.
\end{equation}
Integrating out the momenta $\pi_{\theta}$ and $p_i$ and 
exploiting the corresponding equations of motion and the 
constraints $\ti{\Omega}_a$, we can obtain the desired first-class Lagrangian 
\bea
\label{gilag}
L&=&L_{0}+L_{WZ},\nonumber\\
L_{WZ}&=&(2\dot{q}_{1}-\dot{q}_{2}-2\dot{\theta}+3)\theta
-\frac{1}{2}\dot{\theta}^2,
\eea
where $L_{0}$ is given by Eq. (\ref{action}).  This Lagrangian 
(\ref{gilag}) is invariant under the transformation
\begin{equation}
\delta q_1=-\epsilon_2,~\delta q_2=\epsilon_1,
~\delta q_3=\epsilon_1+2\epsilon_2,~\delta\theta =-\epsilon_2,
\la{trfm1}
\end{equation}
with $\dot{\epsilon}^{1}=\epsilon_{1}+4\epsilon_{2}$ and 
$\dot{\epsilon}^{2}=-\epsilon_{1}-2\epsilon_{2}$, which is obtained from the 
definition of $\delta c \equiv\{c,Q\}$ with the symmetry generator 
$Q=\epsilon_a\tilde{\Omega}_a$.  On the other hand, one can easily see that 
the canonical momenta (\ref{momenta}) are modified in this first-class 
system as 
\beq
\begin{array}{ll}
p_1=\dot{q}_1+2\theta, 
& p_2=\frac{1}{2}(\dot{q}_{3}-\dot{q}_{2})+q_{1}+q_{3}-\theta,\\
p_3=\frac{1}{2}(\dot{q}_{2}-\dot{q}_{3}),
& \pi_\theta=-\dot\theta-2\theta,
\end{array}
\la{ppis}
\eeq
to yield the modified nonholonomic constraint $\ti{\Omega}_{2}$ in terms of 
the $q_{i}$, $\dot{q}_{i}$ and $\dot{\theta}$,
\beq
\ti{\Omega}_{2}=-\dot{q}_{1}+\dot{q}_{2}-\dot{q}_{3}-2q_{3}-1+\dot{\theta},
\la{nhconst1}
\eeq 
which reduces into Eq. (\ref{nhconst0}) in the vanishing limit of the 
auxiliary field $\theta$.  

\subsection{BRST invariant NHCS}

In this section we will obtain the BRST invariant Lagrangian in the framework
of the Batalin-Fradkin-Vilkovisky formalism~\cite{fradkin75} which 
is applicable to theories with the first-class constraints by introducing two
canonical sets of ghosts and anti-ghosts together with auxiliary fields
$({\cal C}^{a},\bar{{\cal P}}_{a})$, $({\cal P}^{a},\bar{{\cal C}}_{a})$,
$(N^{a},B_{a})$, $(a=1,2)$ which satisfy the super-Poisson algebra 
$\{{\cal C}^{a},\bar{{\cal P}}_{b}\}=\{{\cal P}^{a},\bar{{\cal C}}_{b}\}
=\{N^{a},B_{b}\}=\delta_{b}^{a}$.\footnote{Here the super-Poisson bracket is 
defined as $\{A,B\}=\frac{\delta A}{\delta q}|_{r}\frac{\delta B}{\delta p}|_{l}
-(-1)^{\eta_{A}\eta_{B}}\frac{\delta B}{\delta q}|_{r}\frac{\delta A} {%
\delta p}|_{l}$ where $\eta_{A}$ denotes the ghost number in $A$ and the 
subscript $r$ and $l$ the right and left derivatives.}

In the NHCS, the nilpotent BRST charge $Q_{B}$, the fermionic
gauge fixing function $\Psi$ and the BRST invariant minimal Hamiltonian $%
H_{m}$ are given by
\beq
Q_{B}={\cal C}^{a}\tilde{\Omega}_{a}+{\cal P}^{a}B_{a},~~~
\Psi=\bar{{\cal C}}_{a}\chi^{a}+\bar{{\cal P}}_{a}N^{a},~~~ 
H_{m}=\tilde{H}^{\prime}-{\cal C}^{1}
\bar{{\cal P}}_{2}
\eeq
which satisfy the relations $\{Q_{B},H_{m}\}=0$, 
$Q_{B}^{2}=\{Q_{B},Q_{B}\}=0$, $\{\{\Psi,Q_{B}\},Q_{B}\}=0$.
The effective quantum Lagrangian is then given with 
$H_{tot}=H_{m}-\{Q_{B},\Psi\}$ as follows 
\begin{equation}
L_{eff}=p_{i}\dot{q}_{i}+\pi_{\theta}\dot{\theta} +B_{a}\dot{N}^{a}
+\bar{{\cal P}}_{a}\dot{{\cal C}}^{a}
+\bar{{\cal C}}_{a} \dot{{\cal P}}^{a}-H_{tot}.
\end{equation}

Now we choose the unitary gauge $\chi^{1}=\Omega_{1}$, $\chi^{2}=\Omega_{2}$ 
and perform the path integration over the fields $B_{1}$, $N^{1}$, 
$\bar{{\cal C}}_{1}$, ${\cal P}^{1}$, $\bar{{\cal P}}_{1}$ and 
${\cal C}^{1}$ to yield 
\begin{eqnarray}
L_{eff}&=&p_{i}\dot{q}_{i}+\pi_{\theta}\dot{\theta} +B_{2}\dot{N}^{2}
+\bar{{\cal P}}_{2}\dot{{\cal C}}^{2}+\bar{{\cal C}}_{2}\dot{{\cal P}}^{2}  
+\bar{{\cal P}}_{2}{\cal P}^{2}-\pi_{\theta}N^{2}
\nonumber \\
& &-\frac{1}{2}(p^2_1-2p^2_3)-q_1-q_2-q^2_3-(-4p_{3}+4q_{3}-1)\theta
+\frac{1}{2}\pi_{\theta}^{2}\nonumber\\
& &+(2p_{3}-p_{1}-2q_{3}-1)(N^{2}+B_{2}).
\end{eqnarray}
Next, using the variations with respect to $p_{i}$, $\pi_{\theta}$,
${\cal P}$ and $\bar{{\cal P}}$, one obtain the relations
\beq
\begin{array}{ll}
p_1=\dot{q}_1-N^{2}-B_{2}, 
& p_2=\frac{1}{2}(\dot{q}_{3}-\dot{q}_{2})+q_{1}+q_{3}+\theta+N^{2}+B_{2},\\
p_3=\frac{1}{2}(\dot{q}_{2}-\dot{q}_{3})-2\theta-N^{2}-B_{2},
& \pi_\theta=-\dot\theta-N^{2},\\
{\cal P}^{2}=-\dot{{\cal C}}^{2},
&\bar{{\cal P}}_{2}=\dot{\bar{{\cal C}}}_{2},
\end{array}
\la{pisss}
\eeq
to, with the choice of $N^{2}=-B_{2}+2\theta$, yield the effective Lagrangian 
\begin{eqnarray}
L_{eff}&=&L_{0}+L_{WZ}+L_{gh},\nonumber\\
L_{gh}&=&-\frac{1}{2}(B_{2})^{2}-(2\theta+\dot{\theta})B_{2}
+\dot{\bar{{\cal C}}}_{2}\dot{{\cal C}}^{2},
\end{eqnarray}
which is invariant under the BRST transformation
\beq
\begin{array}{lll}
\delta_{B} q_1=-\lambda{\cal C}^{2},
&\delta_{B} q_2=\lambda{\cal C}^{1},
&\delta_{B} q_3=\lambda({\cal C}^{1}+2{\cal C}^{2}),\\
\delta_{B}\theta =-\lambda{\cal C}^{2},
&\delta_{B}\bar{{\cal C}}_{a}=-\lambda B_{a},
&\delta_{B}{\cal C}^{a}=\delta_{B}B_{a}=0,
\end{array}
\la{brsttrfm}
\eeq
with $\dot{\cal C}^{1}={\cal C}^{1}+4{\cal C}^{2}$ and 
$\dot{\cal C}^{2}=-{\cal C}^{1}-2{\cal C}^{2}$, which are generalized 
transformation rules of Eq. (\ref{trfm1}), including the ghost fields.  
Here one notes that the first-class nonholonomic constraint 
$\ti{\Omega}_{2}$ in Eq. (\ref{nhconst1}) is now generalized to include the 
ghost term contributions as follows
\beq
\ti{\Omega}_{2}=-\dot{q}_{1}+\dot{q}_{2}-\dot{q}_{3}-2q_{3}-1
+\dot{\theta}+B_{2},
\la{nhconst2}
\eeq 
and $\ti{\Omega}_{1}$ in Eq. (\ref{effcon}) is trivially satisfied via using 
the relations (\ref{pisss}).

\section{Conclusion}

In conclusion, using the Hamilton-Jacobi (HJ) quantization scheme, we have 
investigated the nonholonomic constrained system, which possesses the 
structure of second-class constraints, to compare with the standard 
and improved Dirac quantization methods (DQMs).  We have shown that the 
integrability conditions in the HJ scheme are equivalent to the involutive 
relations for the first-class constrained system in the improved DQM by 
constructing the first-class Hamiltonian and the corresponding effective 
Lagrangian in the framework of the improved DQM.  Furthermore, with this 
effective Lagrangian, we have also constructed the BRST invariant 
nonholonomic constrained system.  Through further investigation it is 
interesting to apply the improved HJ scheme to the constrained field systems 
as well as constrained point particle ones.

\acknowledgments
We acknowledge financial support from the Korea Research Foundation,
Grant No. KRF-2001-DP0083.

\end{document}